\documentclass[useAMS,usenatbibi,usegraphicx]{mn2e}
\usepackage{psfig}

\newcommand{\be}{\begin{equation}}
\newcommand{\ee}{\end{equation}}
\newcommand{\ba}{\begin{eqnarray}}
\newcommand{\ea}{\end{eqnarray}}

\newcommand{\bml}{\begin{mathletters}}
\newcommand{\eml}{\end{mathletters}}

\newcommand{\etal}{et~al.\ }

\def\kms{\,{\rm km\,s^{-1}}}

\def\ltsima{$\; \buildrel < \over \sim \;$}
\def\simlt{\lower.5ex\hbox{\ltsima}}
\def\gtsima{$\; \buildrel > \over \sim \;$}
\def\simgt{\lower.5ex\hbox{\gtsima}}
\def\gsim{ \lower .75ex \hbox{$\sim$} \llap{\raise .27ex \hbox{$>$}} }
\def\lsim{ \lower .75ex\hbox{$\sim$} \llap{\raise .27ex \hbox{$<$}} }
\def\msun{\,{\rm M_\odot}}

\begin{document}

\title[Hypervelocity Binary Stars]
{Ejection of hypervelocity binary stars by a black hole of intermediate mass orbiting Sgr A$^*$}
\author[A. Sesana et al.]{A. Sesana$^{1}$, P. Madau$^{2}$, \& F. Haardt$^{3\star}$\\
$^{1}$Center for Gravitational Wave Physics, The Pennsylvania State University, University Park, State College, 
PA 16802, USA\\
$^{2}$Department of Astronomy \& Astrophysics, University of California, Santa Cruz, CA 
95064, USA\\
$^{3}$Dipartimento di Fisica e Matematica, Universit\'a dell'Insubria, Via Valleggio 11, 
22100 Como, Italy\\
$^{\star}$Affiliated to INFN}

\date{Received ---}

\maketitle

\begin{abstract}
The discovery of hypervelocity binary stars (HVBs) in the Galactic halo would provide 
definite evidence of the existence of a massive black hole companion  to Sgr A$^*$. 
Here we use an hybrid approach to compute the rate of ejection and the total number of 
HVBs produced by a hypothetical intermediate-mass black hole (IMBH, $M_2<10^5\,\msun$) orbiting
Sgr A$^*$. Depending on the mass of $M_2$ and on the properties of binary stars in the central 
parsec of the Milky Way, we show that the number of undisrupted HVBs expected to be expelled 
from the Galactic Center before binary black hole coalescence ranges from zero to a few 
dozens at most. Therefore, the non-detection of stellar binaries in a complete survey of 
hypervelocity stars would not rule out the occurrence of an IMBH-Sgr A$^*$ in-spiralling 
event within the last few$\times 10^8$ years.
\end{abstract}


\section{Introduction}
Hypervelocity stars (HVSs) are a natural consequence of the presence of a massive black hole (MBH) in
the Galactic Center (GC). At present several HVSs are known to travel in the halo of the 
Milky Way (MW) with Galactic rest-frame velocities between $+400$ and $+750\,\kms$ (Brown et al. 
2005, 2006, 2007). Only the tidal disruption of a tight stellar binary by a single MBH in 
Sgr A$^*$ or the scattering of a single star by a hypothetical MBH binary (MBHB) can kick 
a 3-4 $\msun$ star to such extreme  velocities (e.g. Hills 1988; Yu \& Tremaine 2003, 
hereafter Y03). Direct observational evidence 
for a secondary intermediate-mass hole (IMBH) closely orbiting Sgr A$^*$ is difficult to 
establish, however. In Sesana \etal (2007) we showed that the observed velocity distribution 
of HVSs appears to marginally disfavor the MBHB ejection mechanism, though the statistics 
is still rather poor. Lu \etal (2007, hereafter L07) showed that tight binary stars can be ejected 
by a MBHB without being tidally torn apart: the discovery of just one {\it hypervelocity 
binary star} (HVB) in forthcoming deep stellar surveys could then provide evidence of the 
existence of a massive or intermediate-mass black hole companion to Sgr A$^*$. 

The analysis of L07 is the starting point of this Letter. Our goal is to 
provide an estimate of the number of HVBs expected to be produced by the in-spiral of an 
IMBH onto Sgr A$^*$, using the results of scattering experiments between a MBHB and a bound 
stellar cusp discussed in Sesana \etal (2008, hereafter S08). We will show that a short 
burst of HVSs accompanied by a few HVBs would be an incontrovertible signature of a recent 
in-spiral. By contrast, depending on the properties of the population of binary stars in 
the GC, it is possible that a fast binary black hole in-spiral and coalescence may occur 
without the ejection of a single HVB in the Galactic halo. 

\section{MBHB-star interactions}

\begin{figure}
\centerline{\psfig{file=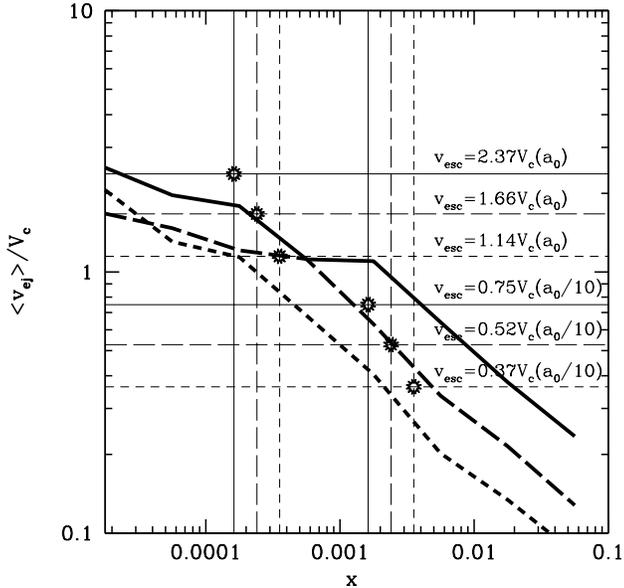,width=86.0mm}}
\caption{{\it Thick curves:} mean stellar ejection velocity (in unit of $V_c$) as a function 
of the dimensionless minimum distance of approach to $M_2$, $x=r_{\rm min,2}/a$. The 
MBHB has an eccentricity of $e=0.1$ and a mass ratio $q=1/81$ ({\it solid line}), 
$q=1/243$ ({\it long-dashed line}) and $q=1/729$ ({\it short-dashed line}). The 
horizontal lines mark the escape velocity $v_{\rm esc}=850\,\kms$ in units of 
$V_c(a)$ for $a=a_0$, $a=0.1\,a_0$, and different mass ratios (using the same line styles 
as above). Similarly, the vertical lines mark the tidal disruption ratio $r_{T,2}/a$ 
for an equal-mass binary with $m_b=2\,\msun$ and $a_b=0.1$ AU (leftmost three lines for $a=a_0$, 
rightmost three for $a=0.1\,a_0$). Dots mark the intersection of corresponding horizontal and
vertical lines dividing the $\langle v_{\rm ej}\rangle/V_c-x$ plane into four quadrants.
To produce an HVB, a thick curve must lie in the corresponding upper right quadrant, where 
$v_{\rm ej}>v_{\rm esc}$ and $r_{\rm min,2}>r_{T,2}$.  Note how, on average, these 
stellar binaries tend to be disrupted and do not become HVBs.
}
\label{fig1}
\end{figure}

Consider a star of mass $m_*$ orbiting the primary hole $M_1$, and assume, for simplicity, 
that the secondary hole $M_2$ $(m_*\ll M_2 \ll M_1)$ is in a circular orbit of radius $a$ 
around $M_1$. When the star experiences a close encounter with $M_2$, its velocity is 
of order the MBHB circular velocity, $v_*\sim V_c=(GM_1/a)^{1/2}$.
A star having closest approach distance to $M_2$ equal to $r_{\rm min,2}\ll a$ 
will be subject to a velocity variation $\Delta v_* \sim (GM_2/r_{\rm min,2})^{1/2}$ as a 
result of a (specific) force $\sim GM_2/r_{\rm min,2}^2$ applied for an encounter timescale 
$\sim (r_{\rm min,2}^3/GM_2)^{1/2}$ (Quinlan 1996; Y03). 
This leads to  
\begin{equation}
{\Delta v_*\over V_c}\sim \left({a M_2\over r_{\rm min,2} M_1}\right)^{1/2}\equiv \sqrt{q/x},
\label{deltav}
\end{equation}
where $x\equiv r_{\rm min,2}/a$ and $q\equiv M_2/M_1$. Since 
in the limit of close energetic encounters the ejection velocity of 
the star is, to first order,  $v_{\rm ej}\propto \Delta v_*$, equation (\ref{deltav})  
shows two important scalings: (1), $v_{\rm ej}$ is inversely proportional to 
the square root of the closest approach distance to $M_2$ during the interaction; and (2),  
the closest approach distance required to eject a star above a given speed (in units of $V_c$) 
scales with the MBHB mass ratio $q$. To verify these simple analytical estimates, we have performed 15 sets 
of 3-body scattering experiments, using the setup described in S08, for mass ratios $q=1/81, 1/243, 
1/729$, and eccentricities $e=0, 0.1, 0.3, 0.6, 0.9$. In each set we integrated 5,000 orbits drawn
from an isotropic distribution of stars bound to $M_1$, and recorded $v_{\rm ej}$, 
$r_{\rm min,2}$, and $r_{\rm min,1}$ (the closest approach distance to $M_1$). 
The stellar semi--major axis $a_*$ is randomly sampled from fifty logarithmic bins spanning
the range $0.03a< a_* < 10a$. The stellar specific angular momentum $L_*$ is
sampled in the interval $[0,L_{*,\rm max}^2]$, according to an equal
probability distribution in $L_*^2$ where $L_{*,\rm max}= \sqrt{GM_1a_*}$ is
the specific angular momentum of a circular orbit of radius
$a_*$. A population of stars with such distribution in
$L_*^2$ has mean eccentricity $\langle e\rangle = 0.66$, 
corresponding to an isotropic stellar distribution 
(e.g. Quinlan, Hernquist, \& Sigurdsson 1995). We stress that, 
on average, the ejection velocity does not depend on the details of the initial Keplerian 
orbit of the star around $M_1$, but only on $r_{\rm min,2}$. Results are plotted in 
Figure~\ref{fig1} for an assumed MBHB eccentricity $e=0.1$. The scaling $v_{\rm ej}\propto x^{-1/2}$ 
breaks down for encounters closer than $x\sim 0.1\,q$: 
the ejection velocity tends to $\langle v_{\rm ej}\rangle\sim 3\,V_c$, the maximum ejection speed
at infinity predicted by simple arguments on elastic scattering. 
We checked that the precise value of $e$ does not play a significant role on determining $\langle v_{\rm ej}\rangle$.  

The third body in our experiments can be thought of 
either as a single star or as a stellar binary. A binary star of 
mass $m_b=m_{*,1}+m_{*,2}$ and semimajor axis $a_b$ is broken apart by tidal forces if 
its center-of-mass approaches a compact object of mass $M$ within the distance (e.g. Miller et al. 2005)
\begin{eqnarray}
r_T&\simeq& \left(3\frac{M}{m_b}\right)^{1/3}a_b\nonumber\\
&\simeq&1.5\times10^{-5}{\rm pc}
\left(\frac{M}{10^4\msun}\frac{\msun}{m_b}\right)^{1/3}
\left(\frac{a_b}{0.1{\rm AU}}\right).
\label{rtd}
\end{eqnarray}
Such ``breakup" radius must be compared with the closest approach distance $r_{\rm min,2}$ 
required for a hypervelocity ejection. If $r_T<r_{\rm min,2}$, a stellar binary may be kicked
to high speeds while preserving its integrity (L07). 

Our three-body approximation does not account for the internal degrees of freedom of the stellar binary. In particular,
a strong interaction with the MBHB can result in the merger of the two stars. Simulations of stellar 
binary-binary interaction in the context of star cluster dynamics show that a merger event is 
a quite common dynamical outcome (Fregeau et al. 2004). However, the dynamical regime 
we consider is different. The stellar binary experiences a complex weak dynamical interaction with the MBHB
(that is unlikely to affect the binary internal structure). Ejection or breakup is caused instead by an instantaneous 
strong encounter with one of the two MBHs. Simulations of strong three body encounters involving a stellar 
binary and a single MBH (with parameters similar to those considered here; Ginsburg \& Loeb 2007) show that
the merger of the two stars happens at most in $\sim$10\% of the cases.      

To make definite predictions, the results of our scattering experiments must be scaled to the GC.
The main parameters of our MW models are summarized in Table~\ref{tab1} (see Sesana et al. 2007 
and S08 for details). The reservoir of stars in the central parsec of the MW is well described by
a power-law density profile, $\rho(r)=\rho_0(r/r_0)^{-\gamma}$, around a $3.5\times10^6\,\msun$ 
MBH.  Here $r_0$ is the characteristic radius within which the total stellar mass is $2M_1$
(the ``radius of influence'' of Sgr A$^*$).  
As the hypothetical secondary hole $M_2$ sinks in, it starts ejecting background stars 
when the total stellar mass enclosed in its orbit is $M_*(<a)\simeq M_2$ (Matsubayashi et al. 2007). 
Following S08, we set the MBHB at initial separation $a_0$ such that $M_*(<a_0)=2M_2$. 
From $\gamma$, $\rho_0$, $r_0$, and $q$ we can derive the parameters $a_0$, $V_{c,0}$, and the 
period $P(a_0)\equiv P_0$. We take a velocity threshold for escaping the MW potential of 
$v_{\rm esc}=850\,\kms$ at $r_0$ (e.g. Smith et al. 2007).  
The horizontal lines in Figure~\ref{fig1} depict the quantity $v_{\rm esc}/V_c$ 
at orbital separation $a=a_0$ and $a=a_0/10$, while vertical lines mark the tidal disruption 
radius $r_{T,2}$ in units of $a$ for the same two MBHB separations. An equal-mass stellar binary
with $m_b=2\,\msun$, and $a_b=0.1$ AU was assumed. The figure shows that, on average, stars must approach 
$M_2$ within a distance $x< r_{T,2}/a$ in order to be ejected. Most binary stars will 
then be tidally disrupted during the strong interaction with $M_2$, and only a few tight binaries 
with $a_b\lsim 0.1$ AU may be survive intact and become HVBs. Note that, while stellar binaries 
can be also broken apart by $M_1$ ($r_{T,1}\gg r_{T,2}$), it is the secondary hole 
$M_2$ that is largely responsible for dissociating candidate HVBs. This is because 
there is no connection between tidal dissociation by $M_1$ and hypervelocity kicks, 
while a close approach to $M_2$, required to gain hypervelocity, can break up the binary. 

 
\begin{table}
\begin{center}
\begin{tabular}{ccccccc}
\hline
$\gamma$ & $r_0$ & $\rho_0$ & $q$ & $a_0$ & $P_0$ & $V_{c,0}$\\
$$ & $[\rm{pc}]$ & $[\rm{\msun pc^{-3}}]$ & $$ & $[\rm{pc}]$ & 
$[\rm{yr}]$ & $[\rm{km s^{-1}}]$\\
\hline
   1.5&    2.25&   7.1$\times10^4$&  1/81&  0.12& 4032& 355\\
   &           &                 &  1/243&  5.8$\times10^{-2}$& 1344& 510\\
   &           &                 &  1/729&  2.8$\times10^{-2}$&  448& 735\\
\hline
\hline
\end{tabular}
\end{center}
\caption{Parameters of the different models. The quantities $\gamma, r_0, \rho_0, q, 
a_0, P_0$, and $V_{c,0}$ are, respectively, the stellar cusp slope, the influence 
radius of Sgr A$^*$, the stellar density at $r_0$, the MBHB mass ratio, its separation when 
ejections start, the MBHB orbital period and circular velocity at $a_0$.
}
\label{tab1}
\end{table}

\section{Hypervelocity stellar binaries}

To quantify the fraction of binary stars that are not disrupted
by $M_2$ (and $M_1$), we need to specify their mass and semi-major distributions.
In our {\it default model}, we assume a log-flat distribution of semi-major axis, 
\begin{equation}
p(a_b)da_b=da_b/a_b,
\label{pflat}
\end{equation}
in the range $10^{-2}<a_b<1$ AU (Heacox 1998). 
The lower limit is set by the contact separation of two solar-mass stars, while the 
upper limit considers that binaries with $a_b>1$AU are unlikely to survive 
the dense stellar environment of the GC (e.g. Y03). We have also run a case with the distribution 
of semi-major axis arising from a log-normal distribution of binary periods $P_b$:
\begin{equation}
p({\rm log}P_b)=C{\rm exp}\left[\frac{-({\rm log}P_b - 
\langle{\rm log}P_b\rangle)^2}
{2\sigma^2_{{\rm log}P_b}}\right].
\label{pnorm}
\end{equation}
Here $C=0.18$ is a normalization constant, $\langle{\rm log}P_b\rangle=4.8$,
$\sigma^2_{{\rm log}P_b}=2.3$, and the period $P_b$ is measured in days (Duquennoy \& Major 1991). 
The above semi-major axis distributions are coupled to two different choices of the stellar 
binary member's mass function (for a total of 4 different models). We either assume all 
binaries to be composed by two equal solar-mass stars ($m_b=2\msun$, {\it default model}), 
or to follow a Salpeter initial mass function (IMF) in the range $1\msun-15\msun$ for $m_{*,1}$ while 
$m_{*,2}$ is randomly chosen in the mass range $1\msun-m_{*,1}$. We shall discuss later the 
effect of these different assumptions on our results. 
 
\begin{figure}
\centerline{\psfig{file=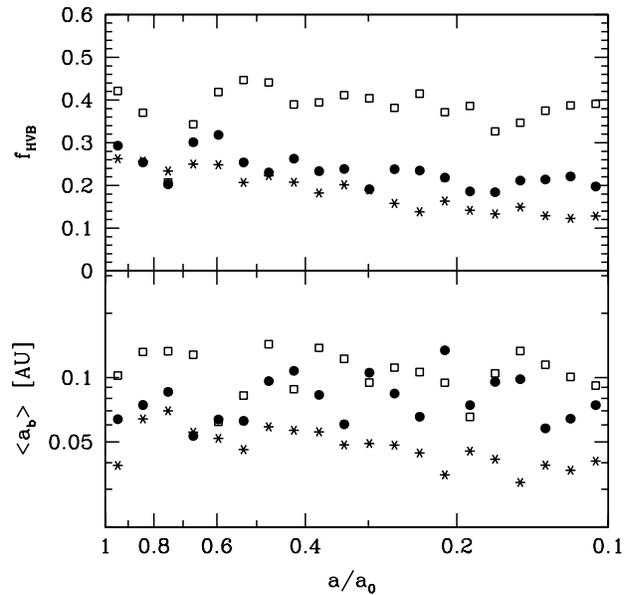,width=86.0mm}}
\caption{{\it Upper panel}: fraction of binaries that survive the interaction with 
the MBHB and are ejected intact with $v_{\rm ej}>v_{\rm esc}$, as a function of MBHB 
separation. {\it Open squares:} $q=1/81$. {\it Filled circles:} $q=1/243$. {\it Stars:} 
$q=1/729$. {\it Lower panel}: mean semi-major axis $a_b$ of undisrupted HVBs. Equal-mass 
binaries with a log-flat distribution in $a_b$ ({\it default model}) have been assumed, and 
the MBHB eccentricity is set to $e=0.6$.}
\label{fig2}
\end{figure}

To estimate the fraction of stellar binaries that survive the interaction with the 
binary black hole and are ejected intact as HVBs, we proceed as follows. For fixed 
$q$ and $e$, we consider orbital separations in the range $0.1-1\,a_0$, select from our 
5,000 simulated orbits those resulting in an ejection with $v_{\rm ej}>v_{\rm esc}$, and denote
their number with $N_{\rm ej}(a)$. We then assume that each of these ``ejection orbits" is 
followed by a binary stellar system with parameters ($a_b,m_{\star,1},m_{\star,2})$ drawn 
from the distributions described above, and calculate the radii $r_{T,2}$ and $r_{T,1}$ 
using equation (\ref{rtd}). Finally, if during the chaotic interaction with the MBH pair
it is $r_{\rm min,2}<r_{T,2}$ or $r_{\rm min,1}<r_{T,1}$, the stellar binary is counted as 
``disrupted before ejection", and added to $N_{\rm TD}(a)$. The fraction of HVBs as a function 
of $a$ is then 
\begin{equation}
f_{\rm HVB}=\frac{N_{\rm ej}(a)-N_{\rm TD}(a)}{N_{\rm ej}(a)}.
\end{equation}
Results are shown in Figure~\ref{fig2} for our default model with MBHB eccentricity $e=0.6$.
 The fraction of undisrupted HVBs is of order $20-40\%$, dropping to $5-20\%$ if the distribution of 
of semi-major axis is derived from equation \ref{pnorm}, similar fractions are 
obtained for all the eccentricity values we sampled. Surviving hypervelocity binaries
have $\langle a_b\rangle\lsim0.1$ AU, i.e. only tight binary stars can be ejected undisrupted.
It is clear from the figure that $f_{\rm HVB}$ and $\langle a_b \rangle$ do not
significantly change as the MBHB shrinks. We can understand this result by noting that 
$\langle v_{\rm ej}\rangle /V_c\propto\sqrt{a/r_{\rm min,2}}$, i.e. $v_{\rm ej}\propto 
r_{\rm min,2}^{-1/2}$. The ejection velocity (in physical units) does not depend then
on MBHB separation, but only on the minimum approach distance to $M_2$.  
Figure~\ref{fig2} also 
shows that the quantities $f_{\rm HVB}$ and $\langle a_b \rangle$ decrease slightly 
with decreasing black hole mass ratios $q$. This occurs because $r_{T,2}\propto M_2^{1/3}
\propto q^{1/3}$, while $r_{\rm min,2} \propto q$, i.e. the more massive the secondary hole 
the weaker the interaction required to kick a star above a given speed. If $q$ is (say) three 
times smaller, a binary star must approach $M_2$ at a distance three times smaller to be 
ejected. But as the breakup radius $r_{T,2}$ decreases by just a factor $3^{1/3}$, fewer tighter 
stellar binaries can survive undisrupted the tidal field of $M_2$.

\section{Ejection rates and detectability}

\begin{figure}
\centerline{\psfig{file=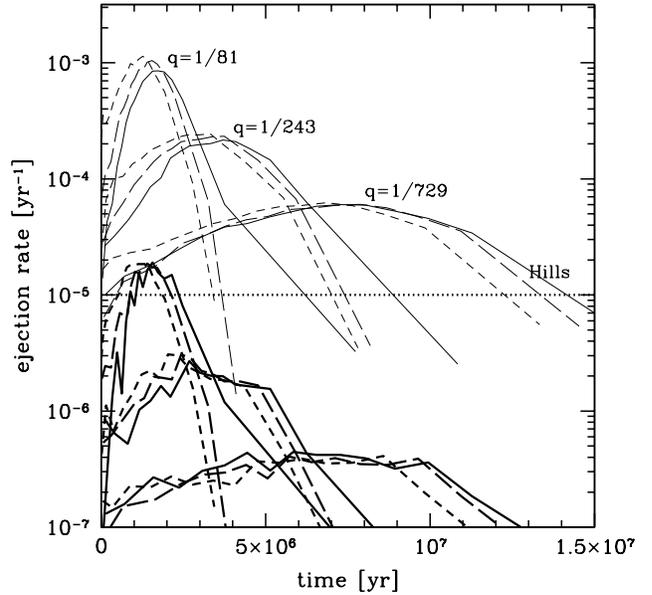,width=86.0mm}}
\caption{Ejection rates of HVSs calculated in S08 ({\it thin curves}) 
versus the ejection rates of HVBs ({\it thick curves}) calculated in this work 
for the default model and $f_b=0.1$. {\it Solid lines:} $e_i=0.1$. {\it Long-dashed 
lines:} $e_i=0.5$. {\it Short-dashed lines:} $e_i=0.9$. The three sets of curves refer, as labeled,
to mass ratios $q=1/81,\,1/243,\,1/729$. The dotted horizontal line marks the ejection rate 
predicted by Hills' mechanism (Y03).} 
\label{fig3}
\end{figure}

We can now estimate the rate at which binary stars would be ejected into the MW halo by 
an IMBH spiralling into Sgr A$^*$. In S08, we self-consistently computed the orbital 
evolution of such an IMBH in terms of $a(t)$ and $e(t)$ (see figure 8 in S08), and estimated the stellar 
mass ejection rate $dm_{\rm ej}/dt$. Here, we assume that a fraction $f_b$ of scattered 
stars are binaries, and account for the evolving MBHB eccentricity during orbital 
decay by linearly interpolating the fraction $f_{\rm HVB}(a,e)$ along the correct $e(a)$ 
curve. The ejection rate of HVBs can be written as
\begin{equation}
R_{\rm HVB}=\frac{1}{\langle m_*\rangle}\frac{dm_{\rm ej}}{dt}
\frac{f_b}{2}f_{\rm HVB},
\label{rate}
\end{equation}
where $\langle m_*\rangle$ is the mean stellar mass.
Results are shown in Figure~\ref{fig3} for our default model and $f_b=0.1$.
The HVB ejection rate peaks between $5\times 10^{-7}-2\times 10^{-5}$ yr$^{-1}$ over 
a timescale of $10^6-10^7$ yr, depending on $q$. For comparison, we also plot the 
ejection rate of HVSs by the in-spiralling IMBH (S08), as well as the rate of HVSs produced
by the tidal disruption of a tight stellar binary by a single MBH in Sgr A$^*$
(Hills' mechanism), as estimated by Y03. In all the cases studied, the total number of HVBs,
$N_{\rm HVB}$ is small compared to the expected number of HVSs. We find $N_{\rm HVB}=28,9,4$ 
for $q=1/81,1/243,1/729$, respectively. If the $a_b$ distribution is log-normal 
(equation~\ref{pnorm}), the fraction of tight binaries is reduced and the number of HVBs 
drops by about a factor of 2. Moreover, in the case of a Salpeter IMF, $\langle m_*\rangle\simeq 
2.5\,\msun$ and $R_{\rm HVB}$ is further reduced by the same factor (see eq. ~\ref{rate}).
The number of HVBs would trivially increase linearly with $f_b$. The number of ejected 
hypervelocity binaries is well approximated by
\begin{equation}
N_{\rm HVB}\approx 280~(170)~ f_{b,<1}~ \frac{\msun}{\langle m_*\rangle} ~ 
\frac{M_2}{5\times10^4\,\msun},
\label{number}
\end{equation}
where $f_{b,<1}$ is the fraction of stars in binaries with $a_b<1$ AU, and 280 (170) is the
normalization constant appropriate for a log-flat (log-normal) $a_b$ distribution.

It should be noted that our approach does not account for the binary stars that are not
initially bound to the MBHB and populate its loss cone because of two-body relaxation 
processes. L07 estimated an HVB ejection rate for such unbound population of few$\times 
10^{-6}$ yr$^{-1}$ in the case of a MBHB with $q=0.01$ and $a=0.0005$ pc. Such a pair 
is expected to have a large eccentricity (e.g. Matsubayashi et al. 2007) and a 
coalescence timescale of only $\sim 10^5$ yr. For larger orbital separations, the loss 
cone is larger but the mean ejection velocity is accordingly smaller, leading to 
lower ejection rates. Such rates are one dex smaller than those we derived for
bound stars and $q=1/81$.   

\section{Summary}

We have applied the hybrid approach described in S08 to compute the rate of ejection 
and the total number of hypervelocity binary stars produced by a hypothetical IMBH
orbiting Sgr A$^*$.  Depending on the mass of $M_2$ and on the properties of binary stars 
in the central parsec of the Milky Way, we have shown that the number of undisrupted HVBs 
expelled before coalescence ranges from zero to a few dozens at most. 
In particular, we have found that the rapid in-spiral of a $5\times10^4\msun$ IMBH 
would generate $\sim 40$ HVBs, assuming a stellar binary fraction of 0.1, 
$m_*=1\,\msun$, and a log-flat distribution of stellar semi-major axis $a_b$. 
A 10\% binary stellar fraction with $a_b<$few AU is suggested by numerical simulations of 
dense stellar clusters (e.g Shara \& Hurley 2006, Portegies-Zwart, McMillan \& Makino 2007).  
The number of HVBs is proportional to the mass of the IMBH and inversely proportional to 
$m_*$, so in the case of a top-heavy stellar mass function (Schodel et al. 2007), the
expected number of HVBs would be lower.  Moreover, if the distribution of stellar
binary semi-major axis is log-normal instead of log-flat, the number of tight binaries 
that can survive a strong interaction with the MBH pair is smaller. The combination of these 
factors can potentially decrease the number of expected HVBs to zero.  

To conclude, while the observation of even a single HVB in the Galactic halo would
be a decisive proof of the recent in-spiralling of an IMBH into Sgr A$^*$, it is likely 
that such an event would give origin to at most a handful of HVBs. Therefore, the 
non-detection of stellar binaries in a complete survey of hypervelocity stars may not be 
used to rule out the existence of an IMBH-Sgr A$^*$ pair in the GC.  

\section*{Acknowledgments}
Support for this work was provided by NASA grant NNG04GK85G (P.M.).



{}

\end{document}